\documentclass[paper,12pt]{JHEP}

\newcommand{\ggN}{g_{YM}^2 N}
\newcommand{\gym}{g_{YM}}

\renewcommand{\r}{\rightarrow}

\newcommand{\gef}{g_{eff}}
\newcommand{\be}{\begin{equation}}
\newcommand{\ee}{\end{equation}}
\newcommand{\eel}[1]{\label{#1}\end{equation}}
\newcommand{\bea}{\begin{eqnarray}}
\newcommand{\eea}{\end{eqnarray}}
\newcommand{\eeal}[1]{\label{#1}\end{eqnarray}}
\newcommand{\baq}{\begin{equation}\begin{array}{rcl}}
\newcommand{\eaq}{\end{array}\end{equation}}
\newcommand{\eaql}[1]{\end{array}\label{#1}\end{equation}}
\newcommand{\beac}{\begin{equation}\begin{array}{rcl}}
\newcommand{\eeacn}[1]{\end{array}\label{#1}\end{equation}}
\newcommand{\ba}{\begin{array}}
\newcommand{\ea}{\end{array}}
\newcommand{\non}{\nonumber \\}
\newcommand{\equ}[1]{(\ref{#1})}
\renewcommand{\a}{\alpha}

\newcommand{\al}{{\alpha^{'}}}
\newcommand{\beq}{\begin{eqnarray}}
\newcommand{\eeq}{\end{eqnarray}}

\newcommand{\w}{Schwarzschild $\:$}

\title{Wilson Loops, Confinement,  and Phase Transitions
 in Large $N$ Gauge Theories from Supergravity}

\author{Andreas Brandhuber, Nissan Itzhaki, 
Jacob Sonnenschein and Shimon Yankielowicz 
 \thanks{Work supported in part by the US-Israel Binational Science
 Foundation, by GIF - the German-Israeli Foundation for Scientific Research,
 and by the Israel Science Foundation.}\\
 School of Physics and Astronomy, \\
 Beverly and Raymond-Sackler Faculty of Exact Sciences, \\
 Tel-Aviv University, Ramat Aviv, Tel-Aviv 69978, Israel\\
 andreasb, sanny, cobi, shimonya@post.tau.ac.il}

\preprint{TAUP-2483-98}

\abstract{We use the recently proposed supergravity
approach to large $N$ gauge theories to calculate ordinary and
spatial Wilson loops of gauge theories in various dimensions.
In this framework we observe an area law for spatial Wilson loops in
four and five dimensional supersymmetric Yang-Mills
at finite temperature. This can be 
interpreted as the area law of ordinary Wilson loops in three and four
dimensional non-supersymmetric gauge theories at zero temperature which
indicates confinement in these theories. 
Furthermore, we show that super Yang Mills theories with 16 
supersymmetries at finite temperature do not admit phase transitions 
between the weakly coupled super Yang Mills and supergravity regimes.
This result is derived by analyzing the entropy and specific heat
of those systems as well as by computing ordinary Wilson loops at 
finite temperature.
The calculation of the entropy was carried out in all different
regimes and indicates that there is no first order phase transition in
these systems.
For the same theories at zero temperature we also compute
the dependence of the quark anti-quark 
potential on the separating distance.}

\keywords{Brane Dynamics in Gauge Theories, Confinement, 
Black Holes in String Theory}
\begin{document}

\section{Introduction}

In the last couple of months it has become clear that supergravity
is a useful tool to study the large $N$ limit of field theories
\cite{mal}. Related work and many new results can be found in 
refs. \cite{hyun} - \cite{ot}.
In the present paper we follow this approach and 
consider phase transitions, Wilson loops
and confinement in supersymmetric and non-supersymmetric field theories
using  supergravity.
We  use the corresponding extremal and near extremal supergravity 
solutions in the 
decoupling limit \cite{mal,juan} and study Wilson loops.
To consider the supersymmetric theories at zero (finite) temperature
we follow \cite{juan1,sjr} (\cite{finite,ste})
and use the Nambu-Goto action
in the background of an extremal (near-extremal) D-brane solution
to calculate the space-time Wilson loop.
We find that for the field theories with maximal supersymmetries there
is no finite temperature phase transition between the SYM and SUGRA
regimes \cite{juan}.
This we also check via a direct entropy consideration which shows that
the entropy matches (up to numerical coefficients which we do not
calculate) across the different domains. Since the entropy seems to
match in all the different regimes of the theory it is most likely
that there is no first order phase transition. This does not exclude
higher order phase transitions which are known to occur in various
cases as we enter the eleven dimensional M theory regime. In those
cases a ``localization" phase transition takes place as discussed in
\cite{juan}.
To consider non-supersymmetric theories at zero temperature we follow
\cite{witten} and consider
the spatial Wilson loop in the background of Euclidean
near-extremal D$p$-brane solutions.
When the spatial size is much larger then $1/T$ (where $T$ is the Hawking
temperature of the near-extremal solution) the effective low energy theory
reduces effectively to a $p$ dimensional non-supersymmetric theory.
Therefore, the spatial Wilson loop gives us  the energy
between a quark and an anti-quark of the $p$ dimensional theory at
zero temperature.
Using this approach we find an area law behavior for
non-supersymmetric YM in three and four dimensions in the large $N$ limit.

The paper is organized as follows:
In section 2 we briefly describe following \cite{juan} the supergravity 
solutions of $N$ coincident D$p$-branes in the field theory 
limit  that was introduced in  \cite{mal}.
In Section 3 we use the supergravity approach to confirm the confinement
behavior of non-supersymmetric YM theory in three (four) dimensions. 
We consider the Euclidean theory of $N$ D3- (D4-) branes with 
$R^3\times S^1$ ($R^4\times S^1$) world volume. In the limit of 
small  radius of the $S^1$ circle,  imposing  appropriate boundary 
conditions, we end up with a three (four) dimensional pure YM theory.  
In this setup we  compute the  spatial Wilson loop 
and  show that it  admits an area law behavior.
Section 4 is devoted to  analyzing the entropy and the specific heat
of the various theories with 16 supersymmetries. It is shown that there is
no phase transition in the SYM, supergravity and other domains of the
D$p$-brane systems. 
In section 5 we present the derivation of ordinary 
Wilson lines (along one space and one time directions) for these
theories at zero temperature.  
This includes the derivation in the ten dimensional supergravity regime,
as well as in the extensions to other energy domains. 
In section 6 the finite temperature behavior of these systems is deduced
from the supergravity description. It is shown that for $p=1 \ldots 4$
they admit a similar behavior as the one discovered in
\cite{finite}, \cite{ste}. 

\section{A very brief review of the theories with 16 supersymmetries}
\label{rev}

In \cite{juan}  
systems of $N$ coincident extremal  Dp-branes  where analyzed 
in the  decoupling  limit \cite{mal}  
\be\label{limit}
 \gym^2=(2\pi)^{p-2} g_s\al^{(p-3)/2}=\mbox{fixed},\;\;~~~~~
\al\r 0,~~~~~~U\equiv \frac{r}{\al}=\mbox{fixed},
\ee
where $g_s=e^{\phi_{\infty}}$, and $\gym $ is the  coupling
constant of the $p+1$ dimensional $U(N)$  SYM theory ( with sixteen 
supercharges) that lives  on the $N$ Dp-branes. 
In the SYM picture $U$ corresponds to finite Higgs expectation value
associated with a  $U(N+1)\r U(N)\times U(1)$ symmetry breaking.
The effective coupling of the SYM theory is $\gef^2=\gym^2NU^{p-3}$.
Perturbation theory can be trusted in the region 
\be
\gef\ll 1.
\ee
The type II supergravity  solution describing $N$ coincident extremal
Dp-branes can be trusted if the curvature in string units and the effective 
string coupling are small.
These conditions yield 
\beq
1 \ll  \gef^2 \ll N^{4 \over 7-p} ~ .
\eeq
which  translate for $p<3$ to the following range of the energy scale $U$
\beq\label{range} 
 (\ggN)^{1/(3-p)} N^{4/(p-3)(7-p)} \ll U\ll (\ggN)^{1/(3-p)} ~.
\eeq 
For $p>3$ the $\ll$ signs are replaced with $\gg$ ones. 
In the supergravity description $U$ is the radial coordinate. 

In general, there are other regions (which will play a role below) 
which take over when the dilaton becomes large.
We do not describe these regions here since they depend on $p$ and
cannot be describes in this concise manner.
For details see \cite{juan}.

\section{Area law }

In this section we use the supergravity approach to field theory and find 
confinement in non-supersymmetric theories in three and four
dimensions.

\subsection{Area law for three dimensional YM theory}

We consider the Wilson loop 
along two space directions in the case of the 
near extremal D3 brane solution.
We will find that it shows area law behavior as in the $R^3$ case in
ref.  \cite{witten}.
We shall take one direction, $Y$, to be large and the other
direction, $L$, to be finite.
In the limit $Y\r\infty$ we  consider configurations
which are invariant under translation in the $Y$ direction.

To describe the theory at finite temperature we go to a Euclidean
description and compactify the time direction on a circle with period
$\beta = T^{-1}$. For large temperature this circle is small and 
the theory becomes effectively the Euclidean description of a three
dimensional field theory with gauge coupling given by
\be
g_{YM_3}^2 = g_{YM_4}^2 T ~.
\ee
Since we choose boundary conditions on the circle such
that supersymmetry is broken, the fermions and scalar fields are
heavy with masses of the order $T$ and $g_{YM4}^2 T$, 
respectively \cite{witten}.
Therefore, at large distances, $L\gg \beta$,
we obtain zero temperature non-supersymmetric
Yang-Mills theory in three dimensions.
Confinement and area law for the Wilson loop are expected. 
We will derive this momentarily by
calculating a spatial Wilson loop of the ``compactified" four
dimensional theory. Since one of the space directions becomes the
Euclidean time direction of the three
dimensional theory this Wilson
line is an order parameter of the theory.
Note that the area law behavior, found in this section,
does not imply confinement in the 3+1 dimensional theory
with temperature.

Confinement is expected to appear in the limit where $L T \gg 1$
i.e. the distance between the quarks is much larger than the size of
the circle. On the other hand if we consider small distances we are
back in the zero temperature 3+1 dimensional theory studied in
\cite{juan1}. Thus, our description interpolates between confinement in
2+1 dimensions ($L T \gg 1$) and Coulomb behavior in ${\cal N}=4$ YM
in 3+1 dimensions 
($L T \ll 1$).

The metric of near extremal D3 branes in the large N limit is 
\bea\label{metric}
&& ds^2  =  \a' \left\{ \frac{U^2}{R^2} 
[ - f(U) dt^2 + dx_i^2] + R^2 f(U)^{-1}
    \frac{dU^2}{U^2} + R^2 d\Omega_5^2 \right\} \non
&& f(U)  =  1 - U^4_T/U^4 \\
&& R^2 = \sqrt{4 \pi g N} ,~~~~~~~~~~~ U_T^4 = 
\frac{2^7}{3} \pi^4 g^2 \mu ~,\nonumber
\eea
where $\mu$ is the energy density.
Thus, the relevant action for the spatial Wilson loop is
\be
S = \frac{Y}{2\pi}\int dx \sqrt{\frac{U^4}{R^4}+\frac{(\partial_x
    U)^2}{1-U_T^4/U^4}}.
\ee
The distance between the quark and the anti-quark is
\be\label{jk}
L = 2 \frac{R^2}{U_0} \int_{1}^{\infty}
\frac{dy}{\sqrt{(y^4-1)(y^4-\lambda)}}, 
\ee
where $\lambda =U_T^4/U_0^4 <1$ and $U_0$ is the minimal value of $U$.
Notice that in the limit $U_0\r U_T$ ($\lambda\r 1$) we get $L\r\infty$.
The energy is 
\be\label{kj}
E=\frac{U_0}{\pi}\int_{1}^{\infty}dy\left( \frac{
    y^4}{\sqrt{(y^4-1)(y^4-\lambda)}}-1\right) +\frac{U_T-U_0}{\pi},
\ee
where, as was explained in \cite{finite}, the subtraction is  at
the horizon.
We are after the large $L$ limit behavior.
Thus we need to take the limit $\lambda\r 1$.
In this limit the main contribution to the integrals in \equ{jk}
and \equ{kj} comes from the region near $y=1$.
Therefore, we get for large $L$
\be
E=T_{QCD} L.
\ee
The tension of the QCD string  is
\be\label{m}
T_{QCD} = \frac{\pi}{2} R^2 T^2,
\ee
where we have used the relation \cite{finite} $U_T=\pi R^2 T$.

It is important to emphasize that the theory for which we obtain
the area law is not YM in three dimensions but ${\cal N}=4 $ at
four dimensions compactified on a circle.
The string tension which we derive, \equ{m},
``knows'' about the four dimensional origin of the theory.
The reason is as follows. 
From \equ{m} we see that the mass of the excitations of the
QCD strings is $M_s= RT$.
The mass associated with the compactification is $M_c=T$.
To trust the supergravity solution we need $R\gg 1$ 
\cite{mal}. Thus $M_s\gg M_c$ and so the QCD strings probe 
distances which are much smaller then the radius of 
compactification.

Some speculations about  string theory at large curvature 
can be made.
For YM in three dimensions we expect to have $T_{QCD} \sim g_3^4$
which in our notation  is $\sim R^8 T^2$
and not $\sim R^2 T^2$ as in \equ{m}.
The system which we are considering here is a
good approximation to YM in three dimensions if
$R \ll 1$ (because then the masses of the 
excitations of the QCD string is lighter then the masses 
associated with the compactification).
The supergravity solution cannot be trusted in that region.
However, since the solution is near-extremal  it makes 
some sense to speculate that the exact string theory solution  
has the same form  but with corrected harmonic functions.
In that case in the limit which we are studying we will get 
for the exact string theory solution an AdS space times a sphere
where the radius of the AdS is not $R^2$ but a function of $R^2$.
The fact that for YM in 3d  we expect to get
$R^8$ for $R^2 \ll 1$ implies that the exact string theory
solution interpolates between $R^2$ at large $R^2$ and $R^8$ at
small $R^2$.

\subsection{Area law in four dimensional YM theory}

The approach of the previous section to confinement
can be generalized to obtain confinement in four 
dimensions from supergravity.
We need to consider the supergravity solution
of near-extremal D4-brane in the decoupling limit.
A D4-brane is described in M-theory as a wrapped M5-brane so
from the point of view of M-theory we relate  the near-extremal
solution of M5-brane to confinement in four dimensions as was 
suggested in sec. 4 of \cite{witten}.

The near-extremal solution of D4-branes in the decoupling limit
is \cite{juan}
\beq\label{nearexD4}
&& ds^2=\a'\left[ \frac{U^{3/2}}{R_4^{3/2}}\left(
(1-U_T^3/U^3)dx_0^2 +dx_1^2+...+dx_4^2\right) +
\right. \non && \left.
 \frac{R_4^{3/2}}{U^{3/2}
(1-U_T^3/U^3)}dU^2 +R_4^{3/2}
\sqrt{U} d\Omega^2_4\right], \\
&& e^{\phi}=\frac{1}{(2\pi)^2} g_5^2 
\left( \frac{U^3}{R^3}\right) ^{1/4},\nonumber
\eeq
where $R_4^{3/2}=g_5 \sqrt{\frac{N}{4\pi}}$ and $g_5$ is the 5D
SYM coupling constant.

We would like to study  the spatial Wilson loop in the 
region where $L T \gg 1$. In this region the effective description
is via a non-supersymmetric YM theory in four dimensions
with coupling constant 
\be\label{54}
\gym^2 =g_5^2 T.
\ee
Unlike the supergravity solution which was used in the previous
section the supergravity solution \equ{nearexD4}, which we use here, cannot
be trusted for arbitrary $U$. $L$,  is related to $U$
by $L\sim 1/U$ and hence it is also bounded.

Before we perform the calculation of the spatial Wilson line
let us first find the upper bound on $L T$ and the bounds on 
 $\gym$ and $\gef^2
 = \gym^2 N$.
The restrictions on $U$ and hence on $U_T$,
are such that the curvature in string units and the effective 
string coupling are small. The result of these restrictions is \equ{range}
\cite{juan} 
\be
\frac{1}{N g_5^2}\ll U_T \ll \frac{N^{1/3}}{g_5^2}.
\ee
Therefore, the supergravity solution can be trusted only for 
distances
\be\label{LL}
N g_5^2\gg L\gg \frac{g_5^2}{N^{1/3}}.
\ee
To find the bound on $T$ we use the relation between $T$ and $U_T$
(the temperature can be obtained from  \equ{nearexD4} and $\left.T
=\frac{1}{4\pi}
\frac{dg_{tt}}{dU}\right|_{U=U_T}$)
\be\label{tem}
T=\frac{3}{2 \sqrt{\pi N} g_5}\sqrt{U_T},
\ee
to get
\be\label{T}
\frac{1}{N g_5^2}\ll T\ll \frac{1}{N^{1/3} g_5^2} ~.
\ee 
From \equ{54} we find that the four dimensional coupling constant is
bounded by
\be
\frac{1}{N} \ll \gym^2 \ll \frac{1}{N^{1/3}} ~.
\ee
We see, therefore, that in the large $N$ limit $\gym $ must go to zero.
For the four dimensional effective coupling, $\gef$, we have
\be
1\ll \gef^2 \ll N^{2/3}.
\ee
Thus the effective four dimensional 
coupling constant
$\gef$ must be large otherwise we cannot trust the 
supergravity description. Finally we turn to the bound for $T L$. 
To be able to  use the supergravity results described below we need 
to find a region where $T L\gg 1$. From \equ{LL} and \equ{T} we get
\be
N^{2/3}\gg TL\gg \frac{1}{N^{4/3}}.
\ee
Therefore, in the large $N$ limit there is a region for which 
we can trust our results. Note that unlike the 3d case, considered 
in the previous section, in the 4d case, for any finite $N$, $L$ is 
bounded.  

Let us now derive the area law behavior.
The action for the string in this case is
\be
S=\frac{Y}{2\pi}\int dx \sqrt{\frac{U^3}{R_4^3}+
\frac{\partial_x U^2}{1-U_T^3/U^3}},
\ee
Using the same manipulations as in \cite{juan1} we get
\beq
&& L=2\frac{R_4^{3/2}}{U_0^{1/2}} \int_{1}^{\infty}
\frac{dy}{\sqrt{(y^3-1)(y^3-\lambda_4)}},\non
&& E=\frac{U_0}{\pi}\int_{1}^{\infty} dy \left( \frac{y^3}{\sqrt{
(y^3-1)(y^3-
\lambda_4)}} - 1 \right) + \frac{U_T - U_0}{\pi},
\eeq
where $\lambda_4=\frac{U_T^3}{U_0^3}$.
For $TL\gg 1$ we have $(U_T-U_0)/U_0\ll 1$ and the integrals are 
dominated by the region close to $y=1$.
Therefore, as in the previous section, we get
\be
E=T_{YM} L
\ee
where the string tension is\footnote{This expression differs by a factor
of $\pi^3$ from the original version of this paper.
We thank C.G. Callan, A. G\"uijosa, K.G. Savvidy and \O. Tafjord for 
pointing out the wrong numerical coefficient.
Also equ. (3.12) had
to be corrected, a factor of $\frac{1}{2 \pi}$ was missing.} 
\be 
T_{YM}=\frac{8 \pi}{27}\gym^2 N T^2 ~.
\ee
This agrees with known large $N$ results if $T$ is identified with
$\Lambda_{QCD}$ up to a $N$ independent constant factor.
Again, as in the previous section, in the region 
where we can trust the supergravity solution, the QCD string 
can probe distances which are much smaller than
the compactification radius and hence it  ``knows''
about the higher  dimensional origin of the underlying theory
(the six dimensional (0,2) theory). 

\subsection{'t Hooft line}

In this section we calculate the energy between a monopole and 
an anti-monopole in the non-supersymmetric four dimensional theory obtained 
from the six dimensional (0,2) upon compactification.
A related discussion for supersymmetric theories can be found in 
\cite{miao}.
The supergravity background is the same as in the previous section.
Namely, it is given by \equ{nearexD4}.
As was explained in the previous section at large distances the 
effective theory is four dimensional along $x_1, x_2, x_3, x_4$.
This theory being related to YM should contain monopoles.
The string theory realization of the monopole is of a D2-brane ending on the
D4-brane. The D2-brane is wrapped along $x_0$ so from the point of view 
of the four dimensional theory it is a point like object.

The action of a D2-branes is 
\be
S=\frac{1}{(2\pi\a')^{3/2}}\int d\tau d\sigma_1 d\sigma_2 e^{-\phi}
\sqrt{\mbox{det} h},
\ee
where $h$ is the induced metric.

For our D2-brane which is infinite along one direction and winds the $x_0$
direction we get
\be
S=\frac{Y}{\gym^2}\int dx \sqrt{\partial_x U^2+ \frac{U^3-U_T^3}{R_4^3}},
\ee
where we have used \equ{54}.
Note the $1/\gym^2$ factor which is expected for a monopole.

The distance between the monopole and the anti-monopole is 
\be
L = 2 \frac{R_4^{3/2}}{U_0^{1/2}} \sqrt{\epsilon} 
 \int_1^{\infty} \frac{dy}{\sqrt{(y^{3} - 1)(y^{3}
    -1 + \epsilon)}} ~.
\ee 
where $\epsilon = 1 - (U_T/U_0)^3$.
The energy (after subtracting the energy corresponding to a free 
monopole and anti-monopole) is 
\be
E=\frac{2 U_0}{(2\pi)^{3/2}
\gym^2}\left[ \int_{1}^{\infty}dy\left( \frac{\sqrt{y^3
        -1+\epsilon}}{{\sqrt{y^3
        -1}}}\right) -1 \right]    +\frac{2(U_T-U_0)}{(2\pi)^{3/2}}.
\ee
We would like to study the system in the region where $L T \gg 1$.
At that region the energy turns positive which means that 
it is energetically favorable for the system to be in a configuration
of two parallel D2-branes ending on the horizon and wrapping $x_0$
which corresponds to zero energy (after the subtraction). So in
the "YM region" we find no force between the monopole and the 
anti-monopole. Which means that there is a screening of the magnetic 
charge.

\section{Entropy and (No)-phase transitions in theories with maximal
  supersymmetry}

In the conformal cases a phase transition cannot take place
at finite temperature.
The reason is that in conformal theories there is no scale and,
therefore, a finite $T_c$ cannot appear in the theory.
Put differently, when compactifying the Euclidean time 
direction the radius of compactification cannot be considered
as large or small simply because there is no scale in the
theory to compare it with.  
In \cite{witten} SYM on $S^3\times S^1$ (rather than
$R^3 \times S^1$ for a superconformal theory in $3+1$ dimensions)
was studied.
The radius of the  $S^3$ serves as a scale in the theory.
Thus a phase transition is possible for this theory.
Indeed, it was shown in \cite{witten} that a phase transition
occurs at finite temperature. 
The phase transition was manifested through a change in the sign
of the specific heat of the AdS black hole \cite{hawpag}.

In this section we would like to consider the non-conformal 
theories with maximal supersymmetries which were studied 
from the supergravity point of view in \cite{juan}. 
These theories, being non-conformal, contain a scale
which depends on $g_{YM}$ and $N$.
Therefore, a phase transition at finite temperature in these 
theories might occur.
At first sight it seems that phase transitions should  appear
in a natural way for the non-conformal theories.
The reason is that different descriptions of the theories are 
valid in different energy regions \cite{juan}.
However, analyzing the entropy of the systems 
describing the theory in  these different
regions, we do not detect any discontinuity in the 
entropy and the specific heat. 
This is an evidence that no
first order phase transition occurs in these theories.
Two  comments are in order: (i) We cannot exclude possible jumps in the 
numerical prefactor of the entropy which, if it exists, will
show a phase transition.
(ii) We cannot exclude higher order phase transitions between regions other
than the SYM and supergravity regions. Such transitions are
known to occur in various cases as we enter into the eleven
dimensional M-theory regime and are associated with
``localization". An example of this phenomenon is the 2+1
dimensional theory \cite{juan}.

As was reviewed in section 1 there are energy regions which are common 
to  all $p$. These  are 
the SYM region and the ten dimensional supergravity region.
There are other regions whose description depends on $p$
\cite{juan}. We shall first show that first order phase transitions do not 
occur in the common regions.
Then we shall consider the other regions. 

\subsection{SYM $\leftrightarrow$ 10d supergravity}

In this section we consider the SYM region, the 10d supergravity region
and the transition between the two regions. Similar results using
a slightly different language have been found in \cite{horopol}.

Perturbation theory in SYM is valid as long as $\gef\ll 1$.
In this regime the interactions between the gluons can be neglected
and the free gas approximation can be used.
For a free gas in $d+1$ dimensions we have
\be\label{gas}
E\sim nVT^{d+1},~~~~~S\sim nVT^d,
\ee
where $n$ is the number of massless fields.
So in our case $n\sim N^2$ and hence  we get
\be\label{ym}
S_{YM}\sim N^{2/(p+1)} V^{1/(p+1)} E^{p/(p+1)}.
\ee

The supergravity solution can be trusted as long as $1/\gef\ll 1$.
(There is also a lower bound which we shall introduce in the  next 
subsections).
The thermodynamics of the system in this region is defined
by the area of the horizon of the supergravity solution which is
 \cite{juan},
\be\label{sg}
S_{sg}\sim\gym^{\frac{p-3}{7-p}} \sqrt{N} E^{\frac{9-p}{2(7-p)}}
V^{\frac{5-p}{2(7-p)}}.
\ee
First we would like to see whether the two entropies 
\equ{ym}, 
\equ{sg} are of the same order at  the transition region,
$\gef\sim 1$.
Since $\gef^2=\gym^2 N U^{p-3}$ the transition is at $U\sim (\gym^2
 N)^{1/(3-p)}$.
The relation between $U$ and $E, V$ is \cite{juan} $U^{7-p}\sim\gym^4
E/V$.
Therefore, the transition should take place at
\be
E=V N^{\frac{7-p}{3-p}} \gym^{-2\frac{p+1}{p-3}}.
\ee
Indeed one can check that precisely at that point
one gets
\be
S_{YM}\sim S_{sg}\sim V \gym^{-2\frac{p}{p-3}} N^{\frac{p-6}{p-3}}.
\ee
Let us consider now the specific heat
$ \left.c=\frac{\partial E}{\partial T}\right|_V $.
From \equ{ym} and \equ{sg} we get
\be\label{spe}
c_{YM}\sim S_{YM},~~~~~c_{sg} \sim S_{sg},
\ee
and hence at the transition $c_{YM}\sim c_{sg}$.
Moreover, it is clear from \equ{spe} that the specific heat is
positive for any temperature and hence there is no first 
order phase transition in
the SYM and/or 10d supergravity regions for any $p$.

\subsection{D1-brane}

Eq.(\ref{sg}) with $p=1$ cannot be trusted all the way to the IR limit
($U\r 0$).
For $U<\gym$ the proper description is via orbifold conformal field
theory \cite{dvv}.
Before we discuss that region we should note that for $U\sim\gym
N^{1/6}$ the correct description is by the S-dual system.
Since the entropy is defined in the Einstein frame and the Einstein
metric is invariant under S-duality the S-dual description yields the
same thermodynamics.

In the region $U<\gym$ the entropy should be calculated in terms of
an orbifold conformal field theory.
Like in the SYM case we consider the entropy in the free theory limit.
The expression we find is a good approximation up to $T\sim
\gym/\sqrt{N}$ \cite{juan} which corresponds to $U\sim\gym$.

The maximal entropy is obtained when the configuration is that of one 
long string whose length is $\tilde{L}=N L$ where $L$ is the 
size of the system \cite{malsus}. 
Therefore, we have 
\be
E\sim  \tilde{L} T^2,~~~~~~~S\sim \tilde{L}T
\ee
which gives
\be
S_{orb}\sim \sqrt{N L E}.
\ee
Since the transition occurs at $U\sim\gym$ and since $U^6\sim \gym^4
E/L$ we find that at the transition region they are of the same order,
\be
S_{orb}\sim S_{sg} \sim \gym\sqrt{N} L.
\ee
Note that $c_{orb} \sim S_{orb}$ and
there is no first order phase transition at
any finite temperature for SYM in $1+1$ dimensions.

\subsection{D2-brane}

Again, like in the previous section,
eq.(\ref{sg}) with $p=2$ cannot be trusted all the way to the IR 
limit.
At the point where \cite{juan}
$U\sim \gym^2$ the correct description becomes the 
conformal theory of coinciding M2 branes with SO(8) R-symmetry.
Therefore, in the region $U<\gym^2$ the entropy is due to the area of
a collection of $N$ near-extremal M2-branes \cite{kletse} 
\be
S_{M2}=\sqrt{N} V^{1/3} E^{2/3}.
\ee
One can easily check \cite{juan} that near the transition region
$U\sim \gym^2$ 
\be
S_{M2}\sim S_{sg}\sim\sqrt{N} V \gym^4.
\ee
Note that $c_{M2} \sim S_{M2}$ and hence there is no first order phase 
transition that can be detected by the entropy .
However, it was pointed out in \cite{juan} that there is a phase 
transition between a translation (along $x_{11}$) invariant solution 
and a localized  one.
This type of phase transition, very likely to be a higher
order one, does not show itself in the entropy or specific heat which
are continuous.

\subsection{D4-branes}

Eq.(\ref{sg}) with $p=4$ cannot be trusted all the way to the UV 
limit, $U\r\infty$.
The correct description for $U > N^{1/3}/\gym^2$ is via M5-branes wrapped
along $x_{10}$ \cite{juan}.
The entropy is \cite{kletse}
\be\label{m5}
S_{M5}\sim\sqrt{N} V_5^{1/6} E^{5/6}
\ee
where $V_5=V_4 2\pi R_{10}$ is the volume of the M5-brane.
Since $R_{10}=\gym^2/(2\pi)^2$ \equ{m5} is the same as \equ{sg} 
for $p=4$.
We conclude that there is no phase transition for the (0,2)
theory on a circle.
Note that for the six dimensional (0,2) theory
a phase transition at finite temperature cannot occur simply because
the theory is conformal and hence there is no scale.
What we show above is that there is no first order
phase transition even if one introduces a scale via compactification.

\subsection{D5-branes}

In the UV there is a transition from the D5-brane solution to 
the NS-fivebrane solution \cite{juan}. The transition is via S-duality.
Since S-duality does not change the Einstein metric and the entropy 
is one quarter of the area in the Einstein frame the entropy is intact.

\subsection{D6-branes}

The correct description  of D6-branes in the UV region, 
$U > \frac{N}{\gym^{2/3}}$, is via \w black hole 
sitting on the $A_{N-1}$ singularity \cite{juan}.
To calculate the entropy one should calculate the area of the 
horizon taking 
into account the $Z_N$ identifications (which 
add a factor of $1/N$ to the usual \w result)
\be
S\sim \frac{\sqrt{N} E^{3/2} \gym^4}{\sqrt{V}}.
\ee 
This coincides with eq.(\ref{sg}) for $p=6$ and hence there is no 
first order phase transition.

\section{ Wilson loops and (no) phase transitions }

In this section we compute the space-time Wilson loops 
for the field theories with maximal supersymmetries both 
at zero and finite temperature.
In these calculations we see no trace of phase transitions in the
supergravity region. This is in accordance with what was derived in
the previous section. 

In the four dimensional case, considered in \cite{finite},
it was found that there are two regions.
For $TL\ll 1$ we found a Coulomb behavior while for 
$TL\gg 1$ the configuration of two parallel strings which end on the brane 
is energetically favored and, therefore, the force between the quark
anti quark vanishes.   
What we find for the non-conformal cases is exactly the same behavior.
This is surprising because the non-conformal theories contain scales
associated with the dimensionful $\gym$ in the relevant dimension.
This supports, therefore, the conclusion of 
the previous section that there is no phase transition
in field theories with maximal number of supersymmetries
in the supergravity region.

Before we calculate the Wilson loop with temperature 
let us first start with the Wilson line at zero temperature.

\subsection{Wilson lines at zero temperature}

In \cite{juan1,sjr} a stringy prescription for the computation of 
the expectation value of the spatial Wilson line of ${\cal N}=4$
was proposed. This expectation value determines 
the dependence of  the  energy $E$ of a quark anti-quark pair
on  the distance between the quarks $L$.
The goal of this section is to extract this dependence for
non-conformal theories with 16 supercharges.
In fact, one such case, 
the ${\cal N}=8$ $U(N)$ SYM in 2+1 dimensions   
was  analyzed in \cite{juan1}.
It was found that  $E\sim ({\ggN\over L^2})^{1/3}$.
It was  further shown \cite{juan1} that this result, 
which is valid in the supergravity regime,  is glued smoothly  with the 
perturbative  SYM result, which is valid in the UV.
In the very IR there is a third regime where the correct description is in
terms of the conformal theory on coinciding M2 branes, but we will not
attempt to treat this here.  
We address now the generalization 
of this result to other Dp-branes systems.

The worldsheet action that corresponds to the metric $G_{MN}$ of  
Dp-branes takes the form 
\beq\label{actionp}
S &=& \frac{1}{2 \pi\al} \int d\tau d\sigma \sqrt{ det G_{MN} 
\partial_\alpha X^M \partial_\alpha X^M}\non
&=& \frac{T}{2 \pi} \int dx \sqrt{(\partial_x U)^2 + 
U^{7-p}/R_p^{7-p}} ~~.
\eeq 
where  $R_p=(g_{YM}^2 d_pN)^{1/(7-p)}$. 

By repeating the procedure of \cite{juan1} one finds
\be
L = 2R_p(\frac{R_p}{U_0})^{(5-p)/2}  \int_1^{\infty}
\frac{dy}{y^{(7-p)/2}\sqrt{y^{7-p} - 1 }} \sim \gym\sqrt{N}
 U_0^{(p-5)/2}.
\eel{length}
The energy  of a system of a quark anti-quark is  (for $p\ne5$)
\be\label{energyp}
E_{sg} \sim -({\ggN\over L^2})^{1/(5-p)},
\ee
the case of $p=5$ is described separately below.
This equation rests on the validity of the 10d
supergravity description and hence it is valid 
as long as the  range of $U$ we integrate over  is in
the supergravity region defined in  section (\ref{rev}). 
This has a double implication: (i) 
The minimal value of $U$, $U_0$,  has to be greater 
than the lower bound stated in \equ{range}, (ii) since one cannot
integrate up to $U=\infty$,  to ensure a reasonable approximation
one has to demand that $x(U_{ub})-L/2\ll L$, where $U_{ub}$
is the upper bound of $U$. For instance  for $p<3$ 
the  consequence is that \equ{energyp} can be trusted only for 
\be (\ggN)^{1/(3-p)} N^{4/(p-3)(7-p)} \ll U\ll (\ggN)^{1/(3-p)} ~.
\ee
In certain cases discussed below we show that in fact the range
can be extended. 

Next we  show,  as in the $2+1$ case \cite{juan1},
that this 10d supergravity result is of the same order as the
perturbative SYM result at the transition region.

\be\label{yy}
E_{YM}\sim \ggN L^{2-p}.
\ee
The transition occurs at  $U\sim (\ggN)^{2/(3-p)}$.
Since  both energies (eq.\equ{yy} and eq.\equ{energyp})   
and the transition region depend only on $g_{YM}^2 N$, with no 
separate dependence on $N$ and $\gym$, it is guaranteed on 
dimensional grounds, that at the transition, 
\equ{energyp}
\be
E_{sg}\sim E_{YM}\sim \ggN^{\frac{1}{3-p}}.
\ee
To describe the behavior of the various systems in 
other regions of energy  we need to do it for each  $p$
separately.

\subsection{D1-brane}

At first sight,  according to \equ{range}, it seems that
\equ{energyp} with $p=1$ is valid only in
the region  $\gym N^{1/6} \ll U_0\ll \gym \sqrt{N}$.
In fact the domain of validity of \equ{energyp} can be extended 
beyond the lower limit $U_0 < \gym N^{1/6}$
where the dilaton becomes large \cite{juan}.
In the region $\gym \ll U_0 \ll\gym N^{1/6}$
the proper description is via the S-dual system.
Namely, the role of the ``quark-anti-quark'' is played by a D1-string
(instead of a fundamental string) and the background is of a
collection of $N$ fundamental strings (and not $N$ D1-branes).
The action of a D1-string in the background of fundamental strings is
the same as the action of F1-string in the background of
D1-branes.
The reason is that due to the difference between their tension 
the D1-string  action is  that of the   fundamental string
action multiplied  with a 
factor $e^{-\phi}$.
On the other hand the S-dual metric contains a factor of
$e^{\phi}$ so that  
these factors  are canceled and one is left 
with the same expression for the energy of the system. 
Only at very low energies $U < \gym$ \cite{juan} the orbifold conformal 
theory \cite{dvv} takes over.

We conclude, therefore, from the supergravity description, 
that for SYM in $1+1$ dimensions in the region
\be
\frac{1}{\sqrt{\ggN}}\ll L\ll\frac{N}{\sqrt{\ggN}}
\ee
the energy between  a``quark anti-quark'' pair  is
\be
E\sim (\sqrt{\frac{\ggN}{L^2}})^{1/4}.
\ee

\subsection{D4-brane}

Beyond the region where the ten dimensional supergravity is valid, for 
$ U\gg {N^{2/3}\over \ggN}$, the system is described 
by  M5-branes wrapped on $R_{10}$ \cite{juan}.
In the case of M5 branes, which was addressed in \cite{juan1}, the 
role of the ``quarks" is played by M2-branes wrapped on $R_{10}$.
The expression for the  energy deduced from the ``Wilson surface" in
the six dimensional (2,0) theory is \cite{juan1}
\be
\frac{E}{L^{'}}\sim -\frac{N}{L^2},
\ee
where $L^{'}$ is the length of the boundary of the
M2-brane on the M5-branes.
In our case since the M2-brane wrapped the $x_{10}$ direction
$L^{'}=2\pi R_{10}\sim \gym^2$.
Thus we obtain the same result as the 10d supergravity expression
\equ{energyp} yields for $p=4$.
Thus, in the UV region $L<N\gym^2$ the energy is
\be
E\sim -\frac{\gym^2 N}{L^2} ~.
\ee

\subsection{D5-brane}

In the five-brane case there are only two regions.
The SYM region and the 10D supergravity region.
Like in the D1-brane case the 10d supergravity region has to be
divided into two region.
But, again like in the D1-brane case, the two regions,
being related by S-duality, yield the same result.
The solution of $U(x)$ takes for $p=5$ the following form
$U(x)=U_0/cos(x/R_5)$ which implies that $L$ does not depend on $U_0$!
\be
L= (2\pi)^{-3/2} \gym \sqrt{N} \sim \sqrt{\alpha' N}.
\ee
This can also be seen also from \equ{length} with $p=5$.

Let us consider the emerging physical picture. Suppose we start at large
distances where we can trust the SYM calculation. The energy of the
quark anti-quark is $E\sim -\ggN /L^3$. For $L<\gym\sqrt{N}$ we find
that $\gef$ becomes larger then $1$ and the supergravity description
takes over. Then $L$ is fixed to be $\gym\sqrt{N}$ and one cannot
decrease $L$ further. 
The fact that for $L<\gym\sqrt{N}$ there is no classical geodesic
should be interpreted according to 
\cite{witten} as having  a zero value of the Wilson loop which
implies an infinite potential. However, a non-trivial 
dependence of $U_0$ on $L$, and hence a non-zero Wilson loop, may 
emerge from a semi-classical quantization of  the system 
with certain collective coordinates.
\footnote{We thank J. Maldacena for pointing this to us.}
At this point we cannot resist to speculate
that perhaps this ``classical" minimal distance is related to the 
existance of a non-locality scale \cite{sei}.

\section{Wilson lines at finite temperature}

Next we address the large $N$ behavior of the Wilson line of  
the non-conformal theories described in the previous section 
at finite temperature. 
As in the discussion of the $p=3$ case this corresponds to 
a description in terms of a  non-extremal supergravity Dp-brane
\cite{juan} which translates into  the  
following  worldsheet action 
\be\label{iop}
S = \frac{T}{2 \pi} \int dx \sqrt{(\partial_x U)^2 + (U^{7-p} -
  U_T^{7-p})/R^{7-p}} ~~.
\eel{action} 
The determination of the $E$ as a function of $L,T,g_{YM}\sqrt{d_pN}$ 
follows the same steps as those leading to \equ{energyp} subjected to
the bounds on $L$ mentioned above. 

The expression for the energy is 
\be
E=\frac{U_0}{\pi}\left[ \int_{1}^{\infty}dy\left( \frac{\sqrt{y^{7-p}
        -1+\epsilon}}{{\sqrt{y^{7-p}
        -1}}}\right) -1 \right]    +\frac{(U_T-U_0)}{\pi}
\ee
where  $\epsilon = 1 - (U_T/U_0)^{7-p}$. For the length we find
\be
L = 2 R_p^{(7-p)/2} U_0^{(p-5)/2} \sqrt{\epsilon} 
 \int_1^{\infty} \frac{dy}{\sqrt{(y^{7-p} - 1)(y^{7-p}
    -1 + \epsilon)}} ~.
\ee 
At small temperatures where $U_T\ll U_0$ one finds the following
 result
\beq\label{energypst}
E(T,L) = c_0 ({\ggN\over L^2})^{1/(5-p)}
\left[ 1+ c ({T {L^2\over \ggN})^{(7-p)/(5-p)}}) \right] 
\eeq

The computation of the Wilson line for arbitrary temperature was 
performed for the case of $p=3$ in \cite{finite}.  
The results for $1 \le p \le 4$ are similar to that found in
\cite{finite}, for $T L \ll 1$ the quark anti-quark pair is connected
by a string and the energy goes as $E \sim -L^{2/(p-5)}$ just like in
the zero temperature case. For large $T L \gg 1$ a configuration of
two parallel strings going from $U = U_T$ to $U = \infty$ is
energetically favored and the force between the quarks vanishes. 
The length  where the string breaks is $c/T$ where $c$ is
a dimensionless constant of order $1$ (which does not depend on $N$,
like in the 4d case \cite{finite}). 
We conclude that also in the
non-conformal cases we find no phase transitions
in the supergravity regime at finite temperature although 
there is a dimensionful parameter in the theory (in contrast to
the conformal p=3 case \cite{finite}).
Maybe one has to choose a compact space to
find a phase transition in field theories with maximal supersymmetries.
For p=3 this has been done in \cite{witten}
where it was found that the four dimensional theory on 
$S^1 \times R^3$ has no phase transition whereas the
theory $S^1 \times S^3$ 
exhibits a phase transition from confinement at small
temperatures to deconfinement at high temperatures (the $S^1$
denotes the compactified Euclidean time direction with period $\beta
= T^{-1}$).
But since in the non-conformal case the space-time is not  
a direct product of $AdS_{p+1}$ times some compact space and we do not
know the metrics that would correspond to spaces with boundaries of the form
$S^1 \times S^p$ we have to leave the study of compact field
theories to the future. 

The discussion in this section was restricted to the supergravity and
Yang-Mills regimes but one may also try to connect to other regimes
\cite{juan} e.g. in the p=2 case one needs the M2 brane description 
in the IR. 

\section*{Acknowledgments}

We would like to thank N. Weiss for useful discussions and especially
J. Maldacena for a critical  reading of the manuscript.


\begin{thebibliography}{99}

\bibitem{mal} J. Maldacena, ``The Large N Limit of Superconformal
  Field Theories and Supergravity'', \hepth{9710014}.

\bibitem{hyun} S. Hyun, ``U-duality between Three and Higher
  Dimensional Black Holes", \hepth{9704005}.

\bibitem{sfetsos} H.J. Boonstra, B. Peeters and K. Skenderis, 
``Duality and asymptotic geometries'', \plb{411}{1997}{59},
\hepth{9706192}; 
K. Sfetsos and K. Skenderis, ``Microscopic derivation of the
Bekenstein-Hawking entropy formula for non-extremal black holes'', 
\hepth{9711138};
H.J. Boonstra, B. Peeters and K. Skenderis, ``Branes and anti-de Sitter
spacetimes,'' \hepth{9801206}.

\bibitem{kall} P. Claus, R. Kallosh, and A. van Proeyen, ``$M$ Five-brane
And Superconformal $(0,2)$ Tensor Multiplet In Six-Dimensions'',
hep-th/9711161; P. Claus, R. Kallosh, J. Kumar, P. Townsend, and
A. van Proeyen,  ``Conformal Field Theory Of $M2$, $D3$, $M5$, and
$D1$-Branes $+$ $D5$-Branes'', \hepth{9801206}.

\bibitem{kall1} R. Kallosh, J. Kumar and A. Rajaraman, ``Special Conformal
Symmetry Of Worldvolume Actions'', \hepth{9712073}.

\bibitem{ferrara1} S. Ferrara and C. Fronsdal, ``Conformal Maxwell Theory As
A Singleton Field Theory On $AdS(5)$, IIB Three-branes and Duality'',
\hepth{9712239}.

\bibitem{minic} M. Gunaydin and D. Minic, ``Singletons, Doubletons, and $M$
Theory'', \hepth{9802047}.

\bibitem{ferrara2} S. Ferrara and C. Fronsdal, ``Gauge Fields As Composite
Boundary Excitations'', \hepth{9802126}.

\bibitem{horowitz} G. T. Horowitz and H. Ooguri, ``Spectrum Of Large $N$
  Gauge Theory From Supergravity,'' \hepth{9802116}. 

\bibitem{juan} N. Itzhaki, J. M. Maldacena, J. Sonnenschein, and
  S. Yankielowicz, ``Supergravity And The Large $N$ Limite Of Theories
  With Sixteen Supercharges'', \hepth{9802042}.

\bibitem{wit} E. Witten, ``Anti-de Sitter Space And Holography",
  \hepth{9802150}.

\bibitem{kachru} S. Kachru and E. Silverstein, ``4d Conformal Field Theories
And Strings On Orbifolds,'' \hepth{9802183}.

\bibitem{berkooz} M. Berkooz, ``A Supergravity Dual Of A $(1,0)$ Field Theory
In Six Dimensions'', \hepth{9802195}.

\bibitem{larsen} V. Balasubramanian and F. Larsen, ``Near Horizon Geometry
And Black Holes In Four Dimensions'', \hepth{9802198}.

\bibitem{gkp}S. S. Gubser, I. R. Klebanov, and A. M. Polyakov,
  ``Gauge Theory Correlators From Noncritical String Theory'',
  \hepth{9802109}. 

\bibitem{flato} M. Flato and C. Fronsdal, ``Interacting Singletons'',
\hepth{9803013}.

\bibitem{vafa1} A. Lawrence, N. Nekrasov, and C. Vafa, ``On Conformal Theories
In Four Dimensions'', \hepth{9803015}.

\bibitem{vafa2} M. Bershadsky, Z. Kakushadze, and C. Vafa, ``String Expansion
As Large $N$ Expansion Of Gauge Theories'', \hepth{9803076}.

\bibitem{ghk} S. S. Gubser, A. Hashimoto, I. R. Klebanov, and M. Krasnitz,
``Scalar Absorption and the Breaking of the World Volume Conformal 
Invariance'', \hepth{9803023}.

\bibitem{volo} I. Ya. Aref'eva and I. V. Volovich, ``On Large $N$ Conformal
THeories, Field Theories On Anti-de Sitter Space, and Singletons'', 
\hepth{9803028}.

\bibitem{casetel} L. Castellani, A. Ceresole, R. D'Auria, S. Ferrara,
  P. Fr\'e and M. Trigiante, ``$G/H$ $M$-Branes And $AdS_{p+2}$
  Geometries'', \hepth{9803039}. 

\bibitem{zaffaroni} S. Ferrara, C. Fronsdal and A. Zaffaroni, ``On $N=8$
Supergravity On $AdS_5$ And $N=4$ Superconformal Yang-Mills Theory'',
\hepth{9802203}.

\bibitem{oz} O. Aharony, Y. Oz and Z. Yin, \hepth{9803051}.

\bibitem{minwalla} S. Minwalla, ``Particles on $AdS_{4/7}$ And Primary
  Operators On $M_{2/5}$ Brane Worldvolumes'', \hepth{9803053}.

\bibitem{leigh} R. G. Leigh and M. Rozali, ``The Large $N$ Limit Of The
$(2,0)$ Superconformal Field Theory'', \hepth{9803068}.

\bibitem{bershadsky} M. Bershadsky, Z. Kakushadze and C. Vafa,
``String Expansion As Large $N$ Expansion Of Gauge Theories'',
\hepth{9803076}. 

\bibitem{rajaraman} A. Rajaraman, ``Two-Form Fields And The Gauge Theory
Description Of Black Holes'', \hepth{9803082}.

\bibitem{gomis} J. Gomis, ``Anti de Sitter Geometry And Strongly
Coupled Gauge Theories'', \hepth{9803119}.

\bibitem{partouche} S. Ferrara, A. Kehagias, H. Partouche and A. Zaffaroni,
``Membranes And Fivebranes With Lower Supersymmetry And Their
$AdS$ Supergravity Duals'', \hepth{9803109}.

\bibitem{minahan} J. A. Minahan, ``Quark-Monopole Potentials In Large
  $N$ Super Yang-Mills'', \hepth{9803111}.

\bibitem{brandt} F. Brandt, J. Gomis and J. Simon, ``D-string on Near Horizon
  Geometries and Infinite Conformal Symmetry", \hepth{9803196}.

\bibitem{halyo} E. Halyo, ``Supergravity on AdS(5/4) x Hopf fibrations
  and conformal field theories", \hepth{9803193}. 

\bibitem{volov} I.V. Volovich, ``Large N gauge theories and anti-de
  Sitter bag model", \hepth{9803174}. 

\bibitem{andria} L. Andrianopoli and S. Ferrara, ``K-K excitations on
  AdS(5) x S**5 as N=4 'primary' superfields", \hepth{9803171}. 

\bibitem{ot} Y. Oz and J. Terning, ``Orbifolds of AdS(5) x S**5 and
  4-D conformal field theories", \hepth{9803167}.

\bibitem{kt} I. Klebanov and A. Tseytlin, ``Intersecting M-branes as
  Four-Dimensional Black Holes'',  \npb{475}{1996}{164}, 
  \hepth{9604166}.

\bibitem{juan1} J. Maldacena, ``Wilson loops in large $N$ field
  theories'', \hepth{9803002}.

\bibitem{sjr} S.J. Rey and J. Yee, ``Macroscopic Strings as
  Heavy Quarks of Large N Gauge Theory and Anti-de Sitter
    Supergravity'', \hepth{9803001}. 

\bibitem{finite} A. Brandhuber, N. Itzhaki, J. Sonnenschein and
 S. Yankielowicz, ``Wilson Loops in the Large $N$ Limit at Finite
 Temperature'', \hepth{9803137}.

\bibitem{ste} S.J. Rey, S. Theisen and J.T. Yee, ``Wilson-Polyakov loop
at finite temperature in large $N$ gauge theory and anti-de Sitter 
supergravity'', \hepth{9803135}.
 
\bibitem{miao} M. Li, "'t Hooft vortices on D-branes", \hepth{9803252}.

\bibitem{sei} N. Seiberg, ``Why is the Matrix Model Correct?'', 
\prl{79}{1997}{3577}, \hepth{9710009}. 

\bibitem{hawpag} S.W. Hawking and D.N. Page, ``Thermodynamics 
of Black Holes in Anti-de Sitter Space'', \cmp{87}{1983}{577}. 

\bibitem{kletse} I.R. Klebanov and A.A. Tseytlin
``Entropy of Near Extremal Black P-Branes'', \npb{475}{1996}{164}, 
\hepth{9604089}.

\bibitem{malsus} J. Maldacena and L. Susskind, ``D-Branes and fat
    black holes'', \npb{475}{1996}{679}, \hepth{9604042}. 

\bibitem{witten} E. Witten, ``Anti-de Sitter Space, Thermal Phase
    Transition, and Confinement in Gauge Theories'', \hepth{9803131}.

\bibitem{dvv} R.  Dijkgraaf, E. Verlinde and H. Verlinde, 
``Matrix String Theory'', \npb{500}{1997}{61}, 
\hepth{9705029}. 

\bibitem{horopol} G.T. Horowitz and J. Polchinski,
  ``Correspondence Principle for Black Holes and Strings",
  \prd{55}{1997}{6189}, \hepth{9612146}. 

\end{thebibliography}
\end{document}